\documentclass[aps,a4paper,12pt]{revtex4}

\usepackage{graphicx}
\usepackage{amsfonts,amsmath,amssymb}
\usepackage[T1]{fontenc}
\usepackage[utf8]{inputenc}
\usepackage[justification=centering]{caption}
\usepackage{braket}

\begin{document}

\graphicspath{{grfs/}}
\newcommand{\Tr}[0]{\text{Tr}}
\newcommand{\rank}[1]{\text{rank}[{#1}]}
\renewcommand{\figurename}{Fig.}

\preprint{}

\title{Comment on ``Quantum discord through the generalized entropy in bipartite quantum states"}%

\author{G. Bellomo}%
\email{gbellomo@fisica.unlp.edu.ar}%
\affiliation{IFLP-CCT-CONICET, Universidad Nacional de La Plata, C.C. 727, 1900 La Plata, Argentina}%
\author{A.P. Majtey}%
\affiliation{Instituto de Fisica, Universidade Federal do Rio de Janeiro, 21.942-972, Rio de Janeiro (RJ), Brazil}%
\author{A.R. Plastino}%
\affiliation{CeBio y Secretaria de Investigacion, Universidad Nacional
del Noroeste de la Prov. de Buenos Aires - UNNOBA
and CONICET, R. Saenz Pena 456, Junin, Argentina}%
\author{A. Plastino}%
\affiliation{IFLP-CCT-CONICET, Universidad Nacional de La Plata, C.C. 727, 1900 La Plata, Argentina}%

\date{\today}%

 \begin{abstract}
 In [2014 \textit{Eur.J.Phys. D} \textbf{68} 1], Hou, Huang, and Cheng present, using Tsallis'
 entropy, possible generalizations of the quantum discord measure,
 finding original results. As for the mutual informations and discord,
 we show here that these two types of quantifiers can take
 negative values. In the two qubits instance we further determine
 in which regions they are non-negative. Additionally, we study alternative generalizations on the basis of Renyi entropies.
 \end{abstract}

\maketitle

On an interesting recent paper, Hou et al. \cite{Hou14} introduce
generalizations for two quantifiers: mutual information and
quantum discords, which they use for the study of quantum
correlations in two qubits systems. It is conventionally agreed
that the mutual information (MI) quantifies total correlations in
bipartite systems. Given a system described by the state
$\rho^{ab}$, with subsystems  $a$ and $b$,  the MI reads
    \begin{equation} \label{eq:mutual_info}
    I(a:b) := S(\rho^a) + S(\rho^b) - S(\rho^{ab}) \,,
    \end{equation}
where $\rho^a:=\Tr_b{\rho^{ab}}$ y $\rho^b:=\Tr_a{\rho^{ab}}$ are
reduced states associated to our subsystems. $S(\cdot)$ is von
Neumann's entropy for a state  $\sigma$:
    \begin{equation} \label{eq:entropy_vn}
    S(\sigma) := -\Tr{(\sigma\log\sigma)} \,.
    \end{equation}
If one wishes to quantify non-classical correlations, these should
be appropriately discriminated  from the total ones. A possibility
is to compute classical correlations via a classical information
measure  (CI)
    \begin{equation} \label{eq:class_info}
    C^b(a:b) := S(\rho^a) - \min_{\{\Pi_i\}}{\sum_k{p_k S(\rho^a_k)}} \,,
    \end{equation}
where $\{\Pi_i\}$ is a complete projective measure, local in $b$,
and
    \begin{equation} \label{eq:meas_out}
    \rho^a_k := \frac{1}{p_k}\Tr_b{[(I_a\otimes\Pi_k)\rho^{ab}(I_a\otimes\Pi_k)]}
    \end{equation}
is the $a$'s conditional state associated to the  outcome $k$ of
$b$. Further,
    \begin{equation}
    p_k := \Tr{[(I_a\otimes\Pi_k)\rho^{ab}(I_a\otimes\Pi_k)]}
    \end{equation}
is the corresponding probability. $I_a$ is the identity operator
for $a$. Eq.~\eqref{eq:class_info} quantifies the classical
correlations from a $b$-perspective and, analogously, one defines
 $C^a$. Given that  \eqref{eq:mutual_info} and
\eqref{eq:class_info} compute quantum and classical correlations,
respectively, the discord measure is given by  \cite{Zur01}

    \begin{equation} \label{eq:discord_vn}
    D^b(a:b) := I(a:b) - C^b(a:b) \,.
    \end{equation}
Hou et al. generalized these measures replacing von Neumann's
entropy by Tsallis' and Renyi's ones (\cite{Ren61,Gel04,Tsa09},
and references therein). The $\alpha$-Renyi quantifier is
\cite{Ren61}

    \begin{equation} \label{eq:entropy_ry}
    S_\alpha(\sigma) := \frac{\log\Tr\sigma^\alpha}{1-\alpha} \,,
    \end{equation}
while Tsallis' counterpart reads  \cite{Tsa09}

    \begin{equation} \label{eq:entropy_ts}
    S_q(\sigma) := \frac{1-\Tr\sigma^q}{(q-1)\ln2} \,.
    \end{equation}
Both quantifiers converge to von Neumann's in the limit
$\alpha\rightarrow1$ ($q\rightarrow1$). Note that we use always
basis-2 logarithms, which slightly modifies the usual definition
of $S_q$.  Hou et al. replace then  $S$ by $S_\alpha$ or $S_q$ in
 \eqref{eq:mutual_info} and \eqref{eq:class_info}, obtaining
generalized mutual information measure  $I_\alpha$ (RMI) and $I_q$
(TMI). We consequently have generalized classical correlations
($C^b_\alpha$, $C^b_q$) and  discords ($D^b_\alpha$, $D^b_q$).

We show below that these last correlation-quantifiers can take
negative values, refuting what is conjectured by Hou et al.
 Even more, the discord can be different from zero, and even negative,
 for classical states.

\vspace{0.5cm} \textit{Rank-three classical states of two qubits.}
 von Neumann's entropy properties guarantee the positivity of $I$, $C^b$, and
$D^b$. We will see that generalized quantifiers do not, in
general, share such positivity property.

\begin{figure}
\centering
\includegraphics[width=.45\textwidth]{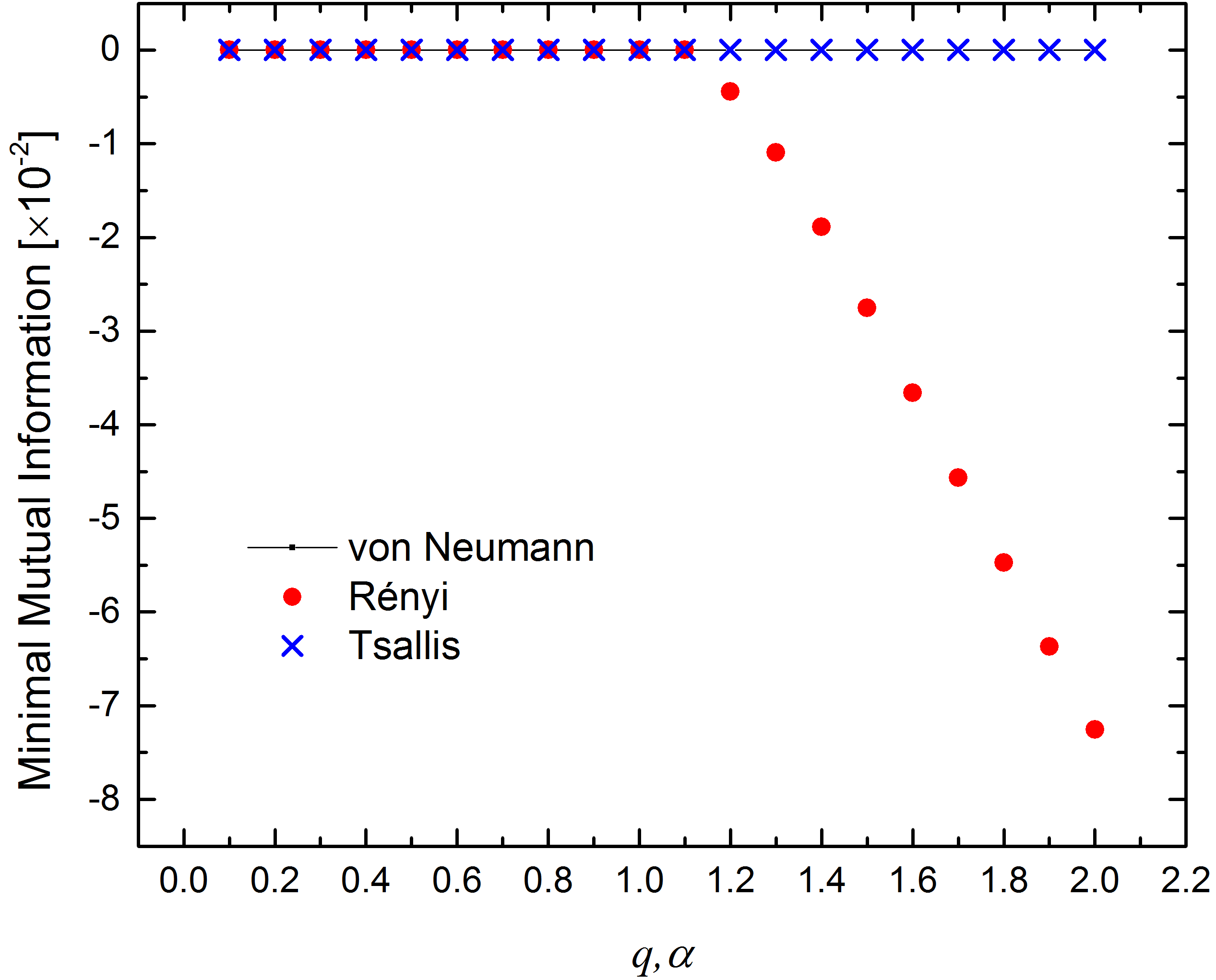}
\includegraphics[width=.45\textwidth]{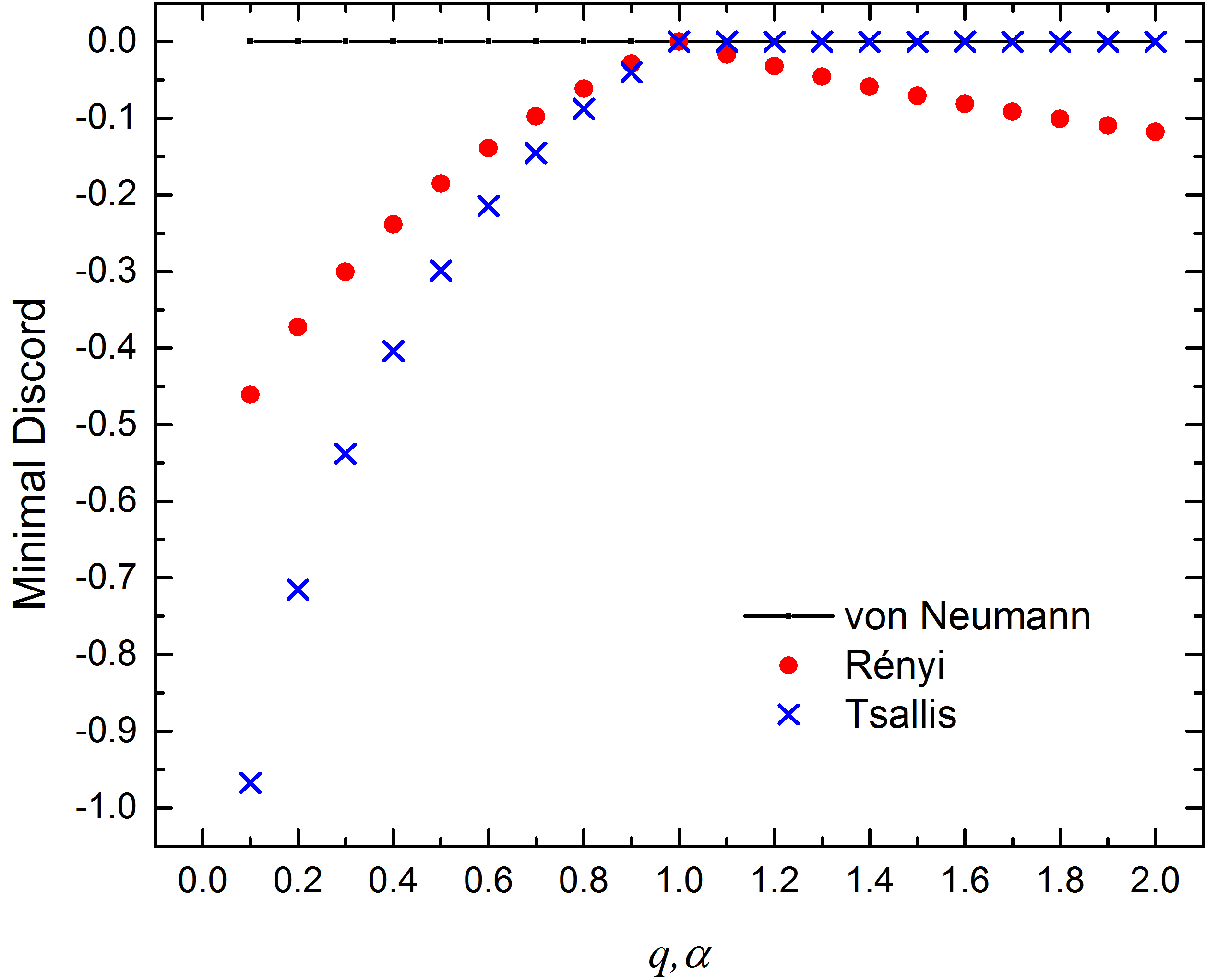}
\caption{Maximum values of the generalized  MI (left) and discord
(right.), for different $\alpha$s andy $q$s.}
\label{fig:min_values}
\end{figure}

\begin{figure}
\centering
\includegraphics[width=.48\textwidth]{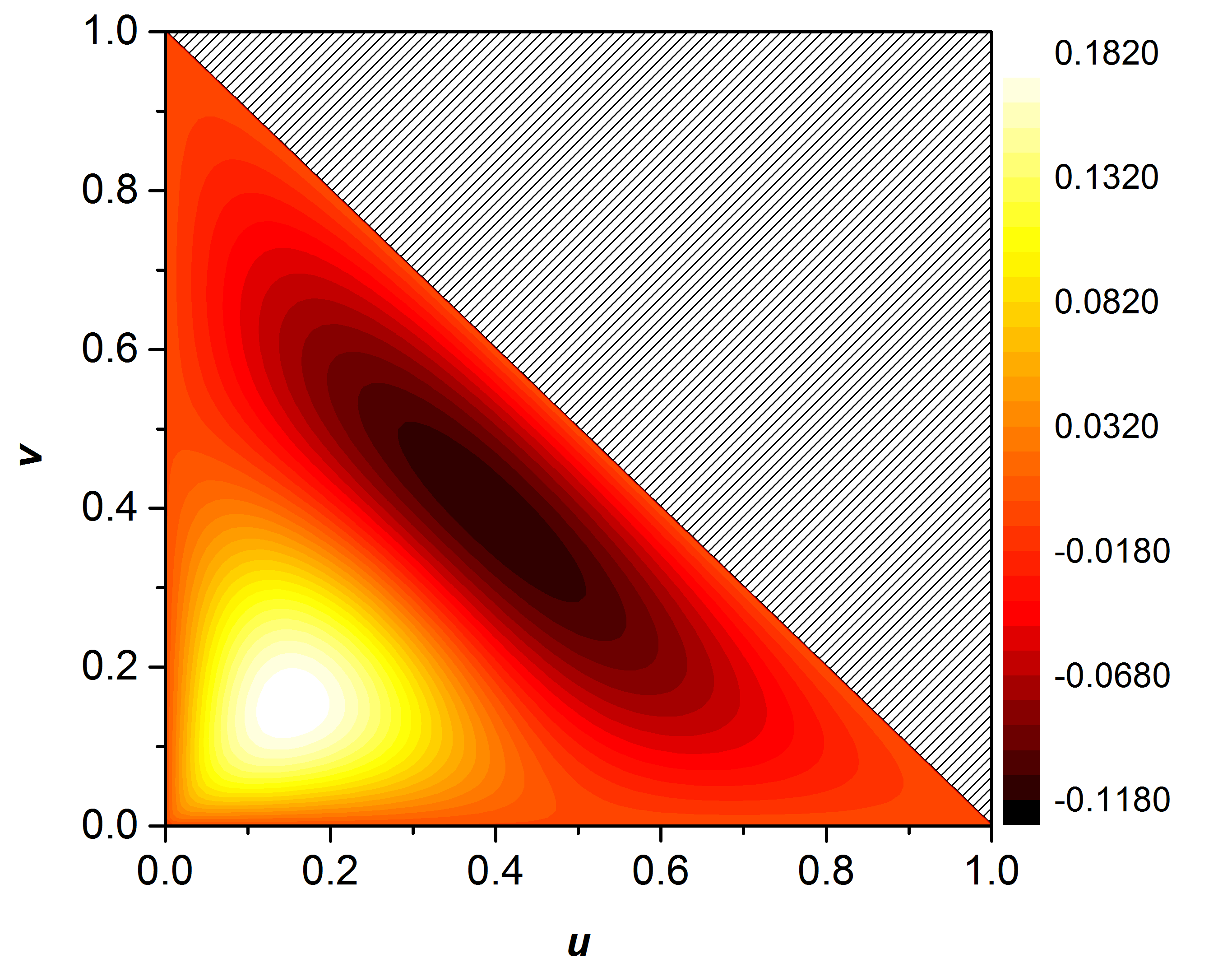}
\includegraphics[width=.48\textwidth]{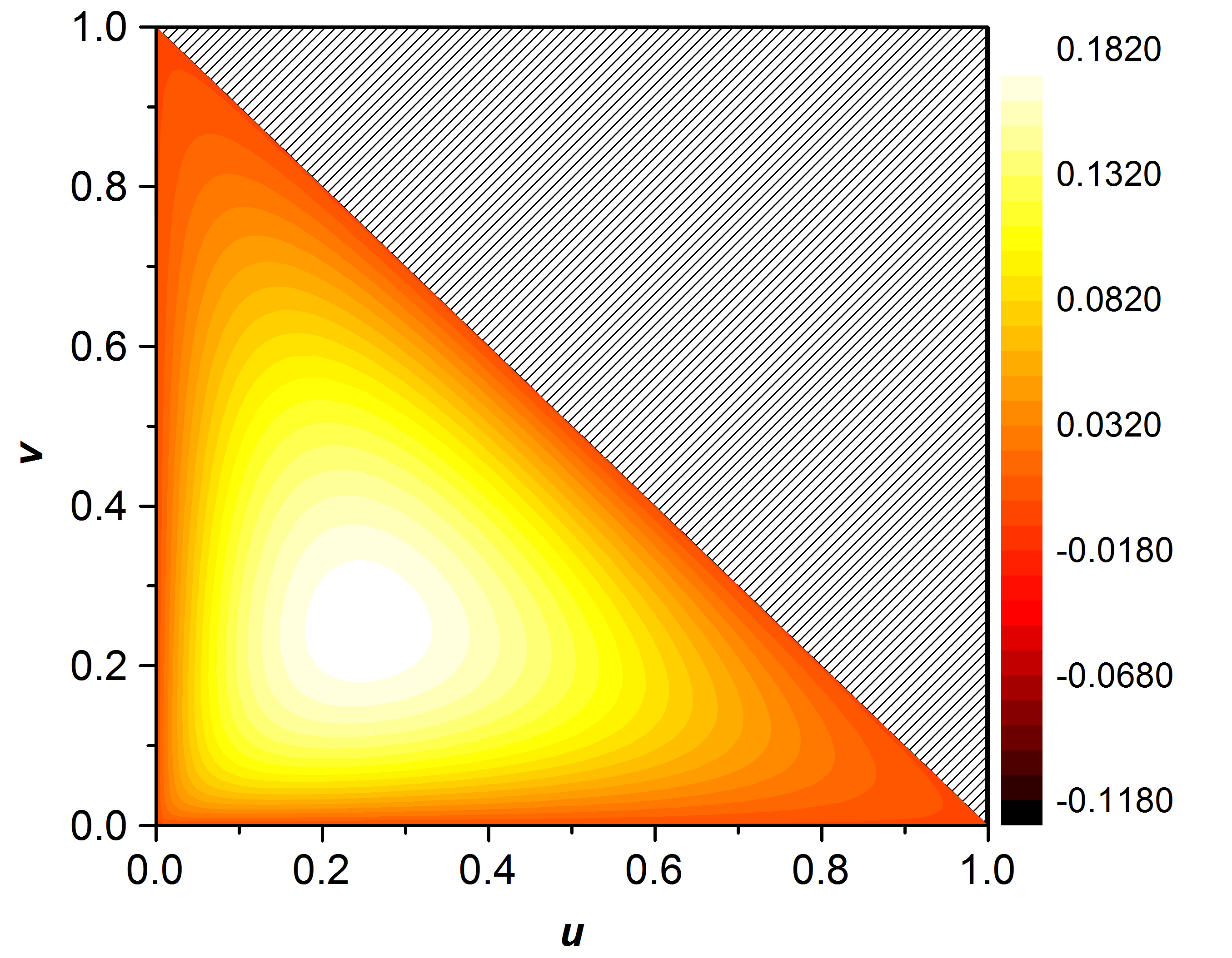}
\caption{Generalized discords for classical states
$\rho^{ab}_{uv}$, with $\alpha=q=2$. Renyi's discord (left) and
Tsallis' one (right). Note the presence of negative values.}
\label{fig:discords_q2}
\end{figure}

As an example consider the family of states given below.
 We focus attention on classical states of range 3 (standard basis).

    \begin{equation} \label{eq:ustate}
    \rho^{ab}_{uv}=
        \left(
        \begin{array}{cccc}
        u & 0 & 0 & 0 \\
        0 & v & 0 & 0 \\
        0 & 0 & 1-u-v & 0 \\
        0 & 0 & 0 & 0
        \end{array}
        \right) \,,
    \end{equation}
with $u,v\geq0$ and $u+v\leq1$. There exists a non-perturbative,
complete and local projective measurement given by the projectors
basis $\{\ket{0}\bra{0},\ket{1}\bra{1}\}$ for the two subsystems.
Thus, for the family  $\rho^{ab}_{uv}$ one has  $D^b(a:b)=0$ (and
$D^a(a:b)=0$). In Figs.
\ref{fig:min_values}--\ref{fig:discords_q2} we note that
generalized measures can be negative \textit{even for classical
states}.Consequently, the ensuing generalized discords cannot
discriminate classical in the sense discussed above.

 CI turns out to be  positive  [all ($\alpha$, $q$)] for the  family
$\rho^{ab}_{uv}$. It would seem that, fora $q>1$, Tsallis' discord
works better, since it is always positive. In the case ($q<1$,
$\alpha<1$), neither Renyi nor Tsallis measures behaves os one
would expect for classical states.

\begin{figure}
\centering
\includegraphics[width=.48\textwidth]{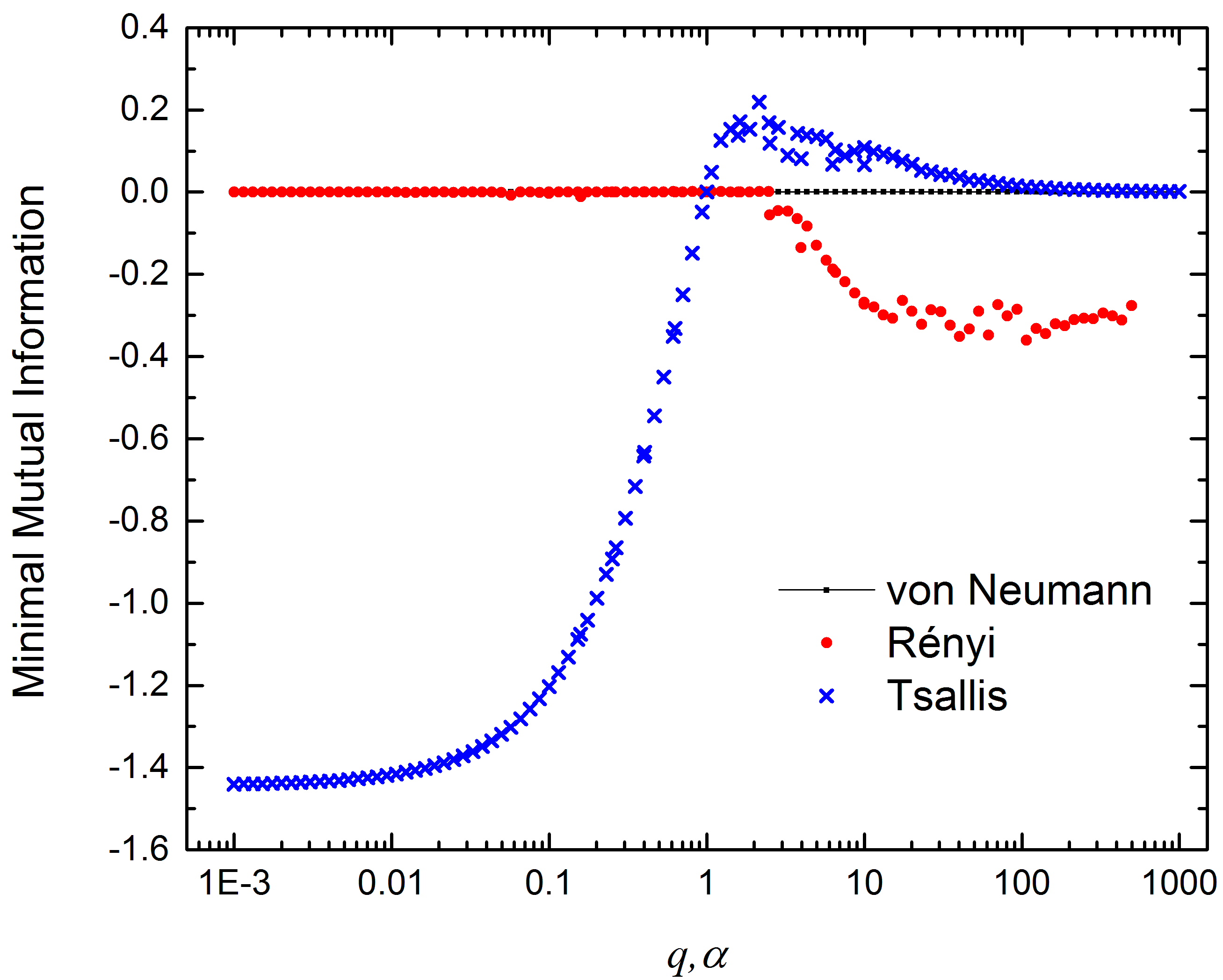}
\includegraphics[width=.48\textwidth]{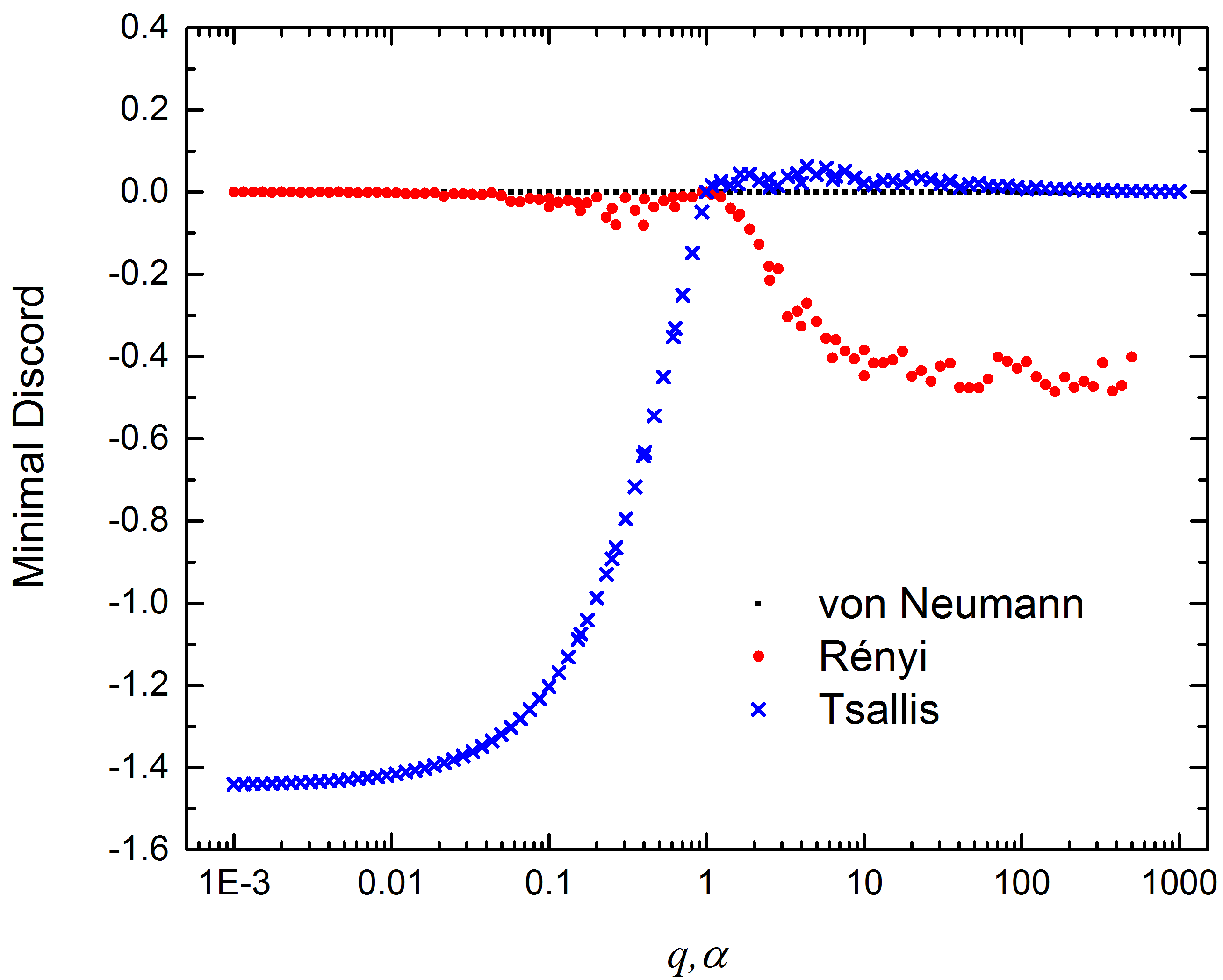}
\caption{Minima for generalized  MI (left) and  discord (right),
using different values of $\alpha$ and $q$, for a large random
sample of states.} \label{fig:rndm_min_values}
\end{figure}

\vspace{0.5cm} \textit{Random states of two qubits.} In this case
we compute  generalized MI, CI, and  discord for different pairs
($\alpha$, $q$) so as to estimate the  range, in such a plane, for
positivity. We considered  $10^5$ random states for each of these
parameters. Fig. \ref{fig:rndm_min_values} plots minima of MI and
discord for a given  $\alpha$ or $q$.

\begin{figure}
\centering
\includegraphics[width=.48\textwidth]{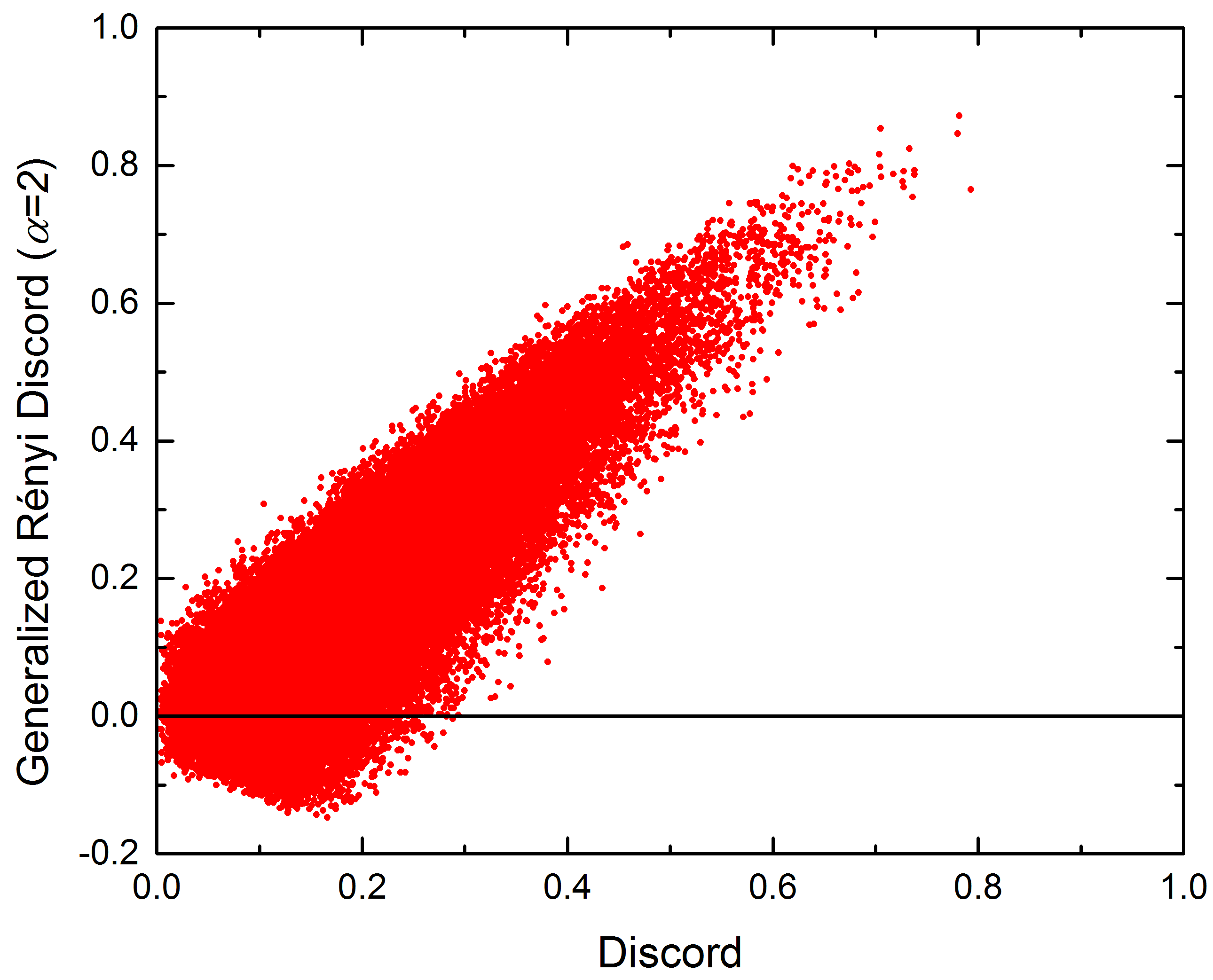}
\caption{Renyi discord with   $\alpha=2$ for  $10^5$ random states
of 2 qubits can be negative for states whose orthodox discord  is
$\lesssim 0.3$.} \label{fig:rndm_dry2_dvn}
\end{figure}

For 2- qubits states, generalized CI's turned out to be positive
for all our states-sample, with  $\alpha$ and $q$    ranging in
$(0,1000)$. This makes it credible that the quantifier is positive
for all ($\alpha$, $q$). Instead, minima for generalized MI and
discord reach negative values for all  $\alpha\neq1$ in the Renyi
instance, while they are positive in the Tsallis case for
$q\geq1$. This would indicate that Tsallis' entropy is strongly
sub-additive (see below). Regretfully enough, the negativity of
these discords does not signal classicality. As an example, states
 that are known to be of a non-classical nature display negative
 Renyi discord for  $\alpha=2$ (Fig.~\ref{fig:rndm_dry2_dvn}).

\vspace{0.5cm} \textit{Alternative generalizations.} It is easy to
see that the von Neumann-\textit{sub-aditivity} (SA) of
$S(\rho^{ab})$:
    \begin{equation} \label{eq:subadditivity}
    S(\rho^{ab}) \leq S(\rho^a) + S(\rho^b) \,,
    \end{equation}
 is tantamount to MI-positivity and that the \textit{concavity}:
    \begin{equation} \label{eq:entropy_convcavity}
    S(\sum_i{p_i\rho_i}) \geq \sum_i{p_iS(\rho_i)} \,,
    \end{equation}
implies that $C^b(a:b)$ and $C^a(a:b)$ are positive measures.
Discord positivity is deduced from  \textit{strong subadditivity}
(SSA) \cite{Zur01,Dat10}:
    \begin{equation} \label{entropy_ssa}
    S(\rho^{abc}) + S(\rho^b) \leq S(\rho^{ab}) + S(\rho^{bc}) \,,
    \end{equation}
equivalent to the  concavity of the conditional entropy
$S(a|b):=S(\rho^{ab})-S(\rho^b)$. In general, generalized
entropies do not share these properties for arbitrary values of
$\alpha$ and $q$. Renyi ones are concave in the interval
$(0,\alpha^*)$, with $\alpha^*=1+\frac{\log4}{\log(N-1)}$,  $N$
being the density matrix range \cite{XuE10,BeZ06}. For
$\alpha\geq\alpha^*$, $S_\alpha$ is neither convex nor concave.
Given $q>1$, Tsallis' entropy is sub-additive so that the
associated mutual information is positive as well, i.e.,
$I_q\geq0$ for $q>1$ \cite{Aud07}. However, for $0<q<1$, Tsallis's
measure is super-additive for product states while for general
states it is neither sub- nor super-additive \cite{Rag95}. Thus,
$I_q$ can adopt negative values for  $0<q<1$. Renyi's entropies
are sub-additive for $\alpha=0$ and $\alpha=1$ \cite{Acz75}. For
all other $\alpha-$values one can find states for which the
associated MI is negative. SSA does not hold in general, save for
the  von Neumann instance \cite{PeV14}. For classical states,
$S_q$ displays SSA if $q\geq1$ \cite{Fur06}. (There exist
particular cases in which $S_\alpha$ also displays  SSA, as, for
instance,  Gaussian states  with $\alpha=2$ \cite{Add12}.) Table
\ref{tab:entropies_prop} details `properties of the different
entropies.

    \begin{table}[h]
    \centering
    \begin{tabular}{c||c|c|c}
                        & Concavity & SA & SSA \\
                        \hline
    $S$                 &\checkmark & \checkmark    & \checkmark \\
    $S_\alpha$          & $(0,1]$   & $\{0,1\}$     & $\times$    \\
    $S_q$               &$(0,\infty)$&  $[1,\infty)$& $\times$
    \end{tabular}
    \caption{Generalized entropies' properties: concavity, sub-additivity (SA), and strong SA (SSA).}
    \label{tab:entropies_prop}
    \end{table}
Concavity and SA are sufficient, but not necessary, to guarantee
positivity. In the case of the range 3-classical family
($\rho^{ab}_{uv}$), our numerical results show that Renyi's CI is
positive for all $\alpha$, being concave only for  $\alpha<3$. As
for discord's positivity, it suffices to demand that
    \begin{equation} \label{eq:fsa}
    I(a:b) \geq \chi(P_a,b) \,,
    \end{equation}
where $\chi(P_a,b):=S(\rho^a)-S(b|P_a)$ is \textit{Holevo's
quantity}  associated to the $b$-state  conditioned to a POVM
measurement of $a$ of operators $P_a$. Coles speaks here of
\textit{firm subadditivity} (FSA), that is less restrictive than
SSA. Hierarchically: SSA$\Rightarrow$FSA$\Rightarrow$SA
\cite{Col11}. Results for a 2 qubits random simulation  (see
Fig.~\ref{fig:rndm_min_values}) would indicate that Tsallis
entropies are  FSA for $q\geq1$, while Renyi ones are FSA for
$\alpha=1$ and, possibly,  for $\alpha=0$.

In von Neumann's entropic scheme, it is equivalent  to define the
MI as the relative entropy between the given state and the product
of the concomitant reduced states, i.e.,
    \begin{equation}\label{eq:mutualinf_relativeent}
    I(a:b):=\min_{\{\sigma^a,\sigma^b\}}S(\rho^{ab}||\sigma^a\otimes\sigma^b) \,,
    \end{equation}
where $S(\rho||\sigma):=-S(\rho)-\Tr(\rho\log\sigma)$  is the
relative entropy, and the minimization runs over the set of all
completely uncorrelated states. Here, Klein's inequality
guarantees the positivity of $I(a:b)$.
Eq.~\eqref{eq:mutualinf_relativeent} offers an
\textit{alternative} path for generalizing the MI in terms of
other entropic measures, different from the one associated to
\eqref{eq:mutual_info}. This alternative was employed by different
authors and is known as the \textit{quantum conditional MI}
\cite{Ber14,Tom09}. Different definitions of Rényi's or Tsallis
relative entropies determine distinct alternatives for the
conditional MI.

A reasonable idea would then entail to define the generalized
mutual information as in Eq.~\eqref{eq:mutualinf_relativeent},
using some generalized relative entropy or divergence:
    \begin{equation} \label{eq:alternative_inf}
    \tilde{I}_\alpha(a:b) := \min_{\{\sigma^a,\sigma^b\}}S_\alpha(\rho^{ab}||\sigma^a\otimes\sigma^b) \,.
    \end{equation}
In similar vein, the classical generalized information will be

    \begin{equation} \label{eq:alternative_class}
    \tilde{C}^a_\alpha(a:b) := {\max_{\{\Pi_i\}}} \min_{\{\sigma^a,\sigma^b\}}S_\alpha({\rho^{ab}}'||\sigma^a\otimes\sigma^b) \,,
    \end{equation}
where
${\rho^{ab}}':=\sum_k{(I_a\otimes\Pi_k)\rho^{ab}(I_a\otimes\Pi_k)}$
is the  posterior state to the measurement of $\{\Pi_i\}$ in $b$.
The new generalized discord would be given by the difference
between these two quantities

    \begin{equation} \label{eq:alternative_dis}
    \tilde{D}^a_\alpha(a:b) := \tilde{I}_\alpha(a:b) - \tilde{C}^a_\alpha(a:b) \,.
    \end{equation}
The positivity of  $\tilde{I}$ and $\tilde{C}^a$ will be
guaranteed by the positivity of the generalized relative
entropies. Noting that our relative entropies fulfill the
\textit{data processing inequality}, $\tilde{D}^a$ will be
positive as well  (see, for instance,  \cite{Mis14}). The scheme
being advanced here should be the subject further exploration.

Recently, the introduction of a new Renyi relative entropy,
monotonous against general quantum (trace preserving) operations,
in the range $1/2\leq q<\infty$ seem to constitute the most
convenient way of computing a states' MI and, a posteriori, to
define a new generalized discord quantifier
\cite{Mul13,Wil13,Lie13,Mis14}.


\begin{thebibliography}{}

\bibitem{Hou14} X.-W. Hou, Z.-P. Huang and S. Chen (2014), Eur. Phys. J. D {\bf 68} (4), 1.

\bibitem{Zur01} H. Ollivier and W.H. Zurek (2001), Phys. Rev. Lett. {\bf 88}, 017901.

\bibitem{Ren61} A. Rényi (1961), {\it On measures of information and entropy}, Proceedings of the fourth Berkeley Symposium on Mathematics, Statistics and Probability 1960, pp. 547-561.

\bibitem{Gel04} M. Gell-Mann and C. Tsallis (Eds.) (2004), {\it Nonextensive Entropy: Interdisciplinary Applications}, Oxford University Press, New York; C. Tsallis (1988), J. Stat. Phys. {\bf 52} 479.

\bibitem{Tsa09} C. Tsallis (2009), {\it Introduction to Nonextensive Statistical Mechanics: Approaching a Complex World}, Springer, New York.

\bibitem{Dat10} A. Datta (2010), arXiv preprint arXiv:1003.5256.

\bibitem{XuE10} D. Xu and D. Erdogmuns (2010), {\it Renyi’s entropy, divergence and their nonparametric estimators}, in {\it Information Theoretic Learning}, p. 47--102, Springer.

\bibitem{BeZ06} I. Bengtsson and K. Zyczowski (2006), {\it Geometry of quantum states: an introduction to quantum entanglement}, Cambridge University Press.

\bibitem{Aud07} Audenaert, K. M. R. (2007), J. Math. Phys. {\bf 48} (8), 083507.

\bibitem{Rag95} G. Raggio (1995), Journal of Mathematical Physics {\bf 36} (9), 4785.

\bibitem{Acz75} J. Aczél and Z. Daróczy, et al. (1975), {\it On measures of information and their characterizations}, Vol. 115 (Academic Press New York).

\bibitem{PeV14} D. Petz and D. Virosztek (2014), arXiv preprint arXiv:1403.7062.

\bibitem{Fur06} S. Furuichi (2006), Journal of Mathematical Physics, {\bf 47} (2), 023302.

\bibitem{Add12} G. Adesso, D. Girolami and A. Serafini (2012), Phys. Rev. Lett. {\bf 109} (19), 190502.

\bibitem{Col11} P.J. Coles (2011), arXiv preprint arXiv:1101.1717.

\bibitem{Ber14} M. Berta, K. Seshadreesan and M.M. Wilde (2014), arXiv preprint arXiv:1403.6102.

\bibitem{Tom09} M. Tomamichel, R. Colbeck and R. Renner (2009), Information Theory, IEEE Transactions on {\bf 55} (12), 5840.

\bibitem{Mis14} A. Misra, A. Biswas, A. Pati, A. Sen De and U. Sen (2014), arXiv preprint arXiv:1406.5065.

\bibitem{Lie13} R.L. Frank and E.H. Lieb (2013), Journal of Mathematical Physics {\bf 54} (12), 122201.

\bibitem{Mul13} M. Müller-Lennert, F. Dupuis, O. Szehr, S. Fehr, and M. Tomamichel (2013), Journal of Mathematical Physics {\bf 54} (12), 122203.

\bibitem{Wil13} M.M. Wilde, A. Winter and D. Yang (2013), arXiv preprint arXiv:1306.1586.


\end{thebibliography}
\end{document}